\documentclass[prl,twocolumn,showpacs,superscriptaddress]{revtex4-1}

\usepackage{amsmath}
\usepackage{amssymb}

\usepackage{longtable}%
\usepackage{dcolumn}
\usepackage{bm}
\usepackage{color}

\usepackage{graphicx}
\usepackage{dcolumn}
\usepackage{bm}
\usepackage{amsmath}
\usepackage{tabu}
\usepackage{multirow}
\usepackage[latin1]{inputenc}
\usepackage{subfigure}
\usepackage{braket}
\usepackage{soul}
\setstcolor{red}
\usepackage{xcolor}
\usepackage{verbatim}
\usepackage{balance}
\usepackage[dvipdfm, pdfstartview=FitH, CJKbookmarks=true, bookmarksnumbered=true,
bookmarksopen=true, colorlinks, pdfborder=001, linkcolor=blue, anchorcolor=blue, citecolor=blue]{hyperref}

\begin{document}
\preprint{Regular article}


\title{Homogeneity range of ternary 11-type chalcogenides Fe$_{1+y}$Te$_{1-x}$Se$_x$}
\author{Cevriye~Koz}
\affiliation{Max Planck Institute for Chemical Physics of Solids,
N\"othnitzer Stra\ss e 40, 01187 Dresden, Germany}
\author{Sahana R\"o{\ss}ler}
\email{roessler@cpfs.mpg.de}
\affiliation{Max Planck Institute for Chemical Physics of Solids,
N\"othnitzer Stra\ss e 40, 01187 Dresden, Germany}
\author{Steffen Wirth}
\affiliation{Max Planck Institute for Chemical Physics of Solids,
N\"othnitzer Stra\ss e 40, 01187 Dresden, Germany}
\author{Ulrich Schwarz}
\affiliation{Max Planck Institute for Chemical Physics of Solids,
N\"othnitzer Stra\ss e 40, 01187 Dresden, Germany}

\date{\today}

\begin{abstract}
The 11-type Fe-chalcogenides belong to the family of Fe-based superconductors. In these compounds, the interstitial Fe is known to strongly influence the magnetic and superconducting properties. Here we present the chemical homogeneity range of ternary compounds 
Fe$_{1+y}$Te$_{1-x}$Se$_x$ based on powder x-ray diffraction, energy dispersive x-ray analysis and magnetization measurements. Our investigations show that the maximum amount of excess Fe in homogeneous Fe$_{1+y}$Te$_{1-x}$Se$_x$ decreases with increase in Se substitution for Te. Using our synthesis procedure, single-phase Fe$_{1+y}$Te$_{1-x}$Se$_x$, with $~0.5 \leq x <$~1 could not be formed for any amount of excess Fe. Further, the superconducting volume fraction in the material is found to be strongly suppressed by excess Fe.
\end{abstract}


\maketitle
%
%


\section*{Introduction}

The 11-type Fe-chalcogenides (Fe-Ch) are considered as representative members of the family of Fe-based superconductors because their crystal structure comprises only of the basic tetrahedral building blocks of edge-sharing Fe(Ch)$_4$ units which are similar to the Fe(As)$_4$ units of the Fe-arsenides (Fe-As). The composition of single-phase material of Fe$_{1+y}$Se with $0 \leq y \leq 0.01$ is very close to stoichiometry \cite{Mc2009,Koz2014}. The superconducting properties of FeSe were found to be extremely sensitive to the amount of excess Fe present in the sample. The superconducting transition temperature $T_c$ decreases drastically with increasing Fe \cite{Mc2009}.  In contrast, the isostructural phase of the heavier homologue tellurium, Fe$_{1+y}$Te, occurs only in the presence of excess Fe ($0.06 \leq y \leq 0.15$) \cite{Rod2011,Ros2011,Koz2013,Rod2013}. The excess Fe is situated in the interstitial $2c$ crystallographic site within the tellurium planes \cite{Bao2009}. Bulk Fe$_{1+y}$Te does not show a superconducting transition, but its magnetic and structural properties can be tuned by changing the amount of excess Fe in the sample \cite{Rod2011,Ros2011,Koz2013,Rod2013,Bao2009}. Substitution of Se for Te in Fe$_{1+y}$Te induces superconductivity with a maximum $T_c \approx$ 15 K  observed for $\approx$ 50 $\%$ Se substitution \cite{Yeh2008,Fang2008,Liu2009,Ros2010,Che2015}. Also for the substituted materials, the superconducting as well as the normal state properties of Fe$_{1+y}$Te$_{1-x}$Se$_x$ are found to be influenced by excess Fe. In the normal state, a charge carrier localization in the electrical transport has been observed in Fe$_{1+y}$Te$_{0.5}$Se$_{0.5}$ for higher Fe concentrations \cite{Liu2009,Ros2010}. Since the concentrations of excess Fe in single phase materials of Fe$_{1+y}$Te \cite{Koz2013} and Fe$_{1+y}$Se \cite{Koz2014} are substantially different, a composition gradient of Fe can be expected in the substitution series of 
Fe$_{1+y}$Te$_{1-x}$Se$_x$. To our knowledge, a careful investigation of the chemical homogeneity range of Fe$_{1+y}$Te$_{1-x}$Se$_x$ is still lacking even though the knowledge of the chemical homogeneity range of these materials is of utmost importance for a proper interpretation of more complex phenomena such as the coexistence of magnetism and superconductivity. Therefore, we synthesized a series of polycrystalline 
Fe$_{1+y}$Te$_{1-x}$Se$_x$ and investigated the properties by powder x-ray diffraction (PXRD), energy dispersive x-ray spectrosocpy (EDX), and magnetization measurements to establish the homogeneity range of the ternary phase. 
\section{Experimental}
\label{sec:1}
Polycrystalline samples of Fe$_{1+y}$Te$_{1-x}$Se$_x$ were synthesized by solid-state reaction. More than sixty compounds with different compositions in the range  $0\leq y \leq 0.15$ and $0\leq x \leq 1$ were synthesized by taking appropriate mixtures of nominal amounts of Fe, Se and Te. Starting materials were heated up to 973 K with a rate of 100 K/h and kept at this temperature for 24 hours before increasing the temperature to 1193 K. The dwelling at 1193 K for 24 h was followed by cooling to 973 K with a rate of 100
K/h (50 K/h), and further annealing for 12 hours. Finally, samples were cooled to room temperature at a rate of 100 K/h. In specific cases, the samples were annealed at 973 K for 48 h to enhance the homogeneity. For the nominal compositions with $x \geq 0.5$, a lower annealing temperature (673 K) was used. All synthesized materials were characterized by PXRD and EDX analysis. The lattice parameters were determined using the diffraction lines of LaB$_{6}$ as an internal standard. 
\section{Results and discussion}
\begin{figure}[t]
  \includegraphics[width=0.475\textwidth]{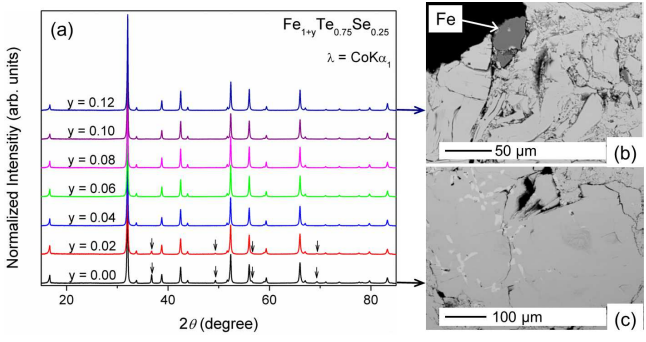}
\caption{(a) Powder x-ray diffraction diagrams (PXRD) of Fe$_{1+y}$Te$_{0.75}$Se$_{0.25}$ ($y$ = 0.00-0.12) annealed at 973 K
for 2 days. Back scattered electron (BSE) images of annealed (b) Fe$_{1.12}$Te$_{0.75}$Se$_{0.25}$ and (c) Fe$_{1.00}$Te$_{0.75}$Se$_{0.25}$. In
PXRD, arrows indicate the Fe-deficient second phase, Fe$_{0.69(1)}$Te$_{0.79(1)}$Se$_{0.21(1)}$. Dark
region in (b) and light regions in (c) correspond to unreacted iron and Fe-deficient
second phase, respectively.}
\label{fig:1}       
\end{figure}

\begin{figure}[t]
  \includegraphics[width=0.475\textwidth]{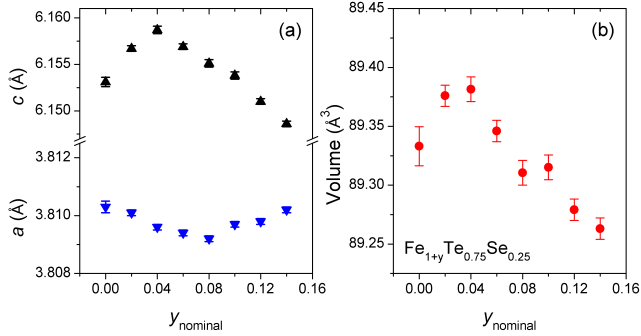}
\caption{(a) Lattice parameters and (b) unit cell volume as a function of 
nominal Fe composition for annealed samples Fe$_{1+y}$Te$_{0.75}$Se$_{0.25}$. For $ y \geq 0.12$
and $ y \leq 0.02$, samples contain unreacted Fe and Fe$_{0.69(1)}$Te$_{0.79(1)}$Se$_{0.21(1)}$, respectively.}
\label{fig:2}       
\end{figure}

The PXRD patterns and back scattered electron (BSE) images of Fe$_{1+y}$Te$_{0.75}$Se$_{0.25}$ for $0\leq y \leq 0.12$ are presented in Fig. \ref{fig:1}. For samples with $y \geq 0.12$, the EDX analysis confirms the presence of unreacted Fe. The BSE image of the sample $x$ = 0.25, $y$ = 0.12 is presented in Fig. \ref{fig:1}(b), in which the elemental Fe is indicated by an arrow. In the case of PXRD, the reflection corresponding to unreacted $\alpha$-Fe overlaps with the main phase and hence, could not be detected. 
%
For low Fe contents ($y \leq 0.02$), a
second phase with EDX composition Fe$_{0.69(1)}$Te$_{0.79(1)}$Se$_{0.21(1)}$ is observed, see Fig. \ref{fig:1}(c),
The peak positions of this second phase suggest that the impurity phase is related to
the structure motif of hexagonal Fe$_{0.67}$Te ($P6_3/mmc$) \cite{Gro1954}. The refined lattice
parameters of the second phase are $a$ = 3.7779(2) \AA~ and $c$ = 5.6668(5)~ \AA. These lattice
parameters are larger than the reported values for NiAs-type Fe$_{0.685}$Te$_{0.8}$Se$_{0.2}$ 
($a$ = 3.771 \AA~and $c$ = 5.660 \AA) \cite{Ter1981} Single phase Fe$_{1+y}$Te$_{0.75}$Se$_{0.25}$ samples can be
obtained for $0.02 < y < 0.12$. Lattice parameters and unit cell volumes as a function of
the Fe content are given in Fig. \ref{fig:2}. With increasing amount of Fe, lattice
parameters and volume decrease within the homogeneity range. The compositions obtained from the EDX analysis are presented in Table \ref{tab:1}.
The EDX results are in agreement with the PXRD analysis.
\begin{table}[t]
\caption{Compositions according to EDX measurements of polycrystalline samples
Fe$_{1+y}$Te$_{0.75}$Se$_{0.25}$ after annealing at 973 K for 2 days. * indicates the presence of a second phase Fe$_{0.69(1)}$Te$_{0.79(1)}$Se$_{0.21(1)}$}
\label{tab:1}       
\centering
\begin{tabular}{lll}
\hline\noalign{\smallskip}
y & Phase 1 & Phase 2\\  
\noalign{\smallskip}\hline\noalign{\smallskip}
0.00 & Fe$_{0.99(2)}$Te$_{0.71(2)}$Se$_{0.29(2)}$ & *\\
0.04 & Fe$_{0.98(1)}$Te$_{0.70(1)}$Se$_{0.30(1)}$ &     \\
0.06 & Fe$_{1.03(1)}$Te$_{0.75(2)}$Se$_{0.25(2)}$ &      \\
0.08 & Fe$_{1.06(2)}$Te$_{0.77(2)}$Se$_{0.23(2)}$ &\\
0.10 & Fe$_{1.10(5)}$Te$_{0.77(1)}$Se$_{0.23(1)}$ &\\
0.12 & Fe$_{1.07(1)}$Te$_{0.72(1)}$Se$_{0.28(1)}$ & Fe\\
\noalign{\smallskip}\hline
\end{tabular}
\end{table}
\begin{figure}[h]
\centering
  \includegraphics[width=0.4\textwidth]{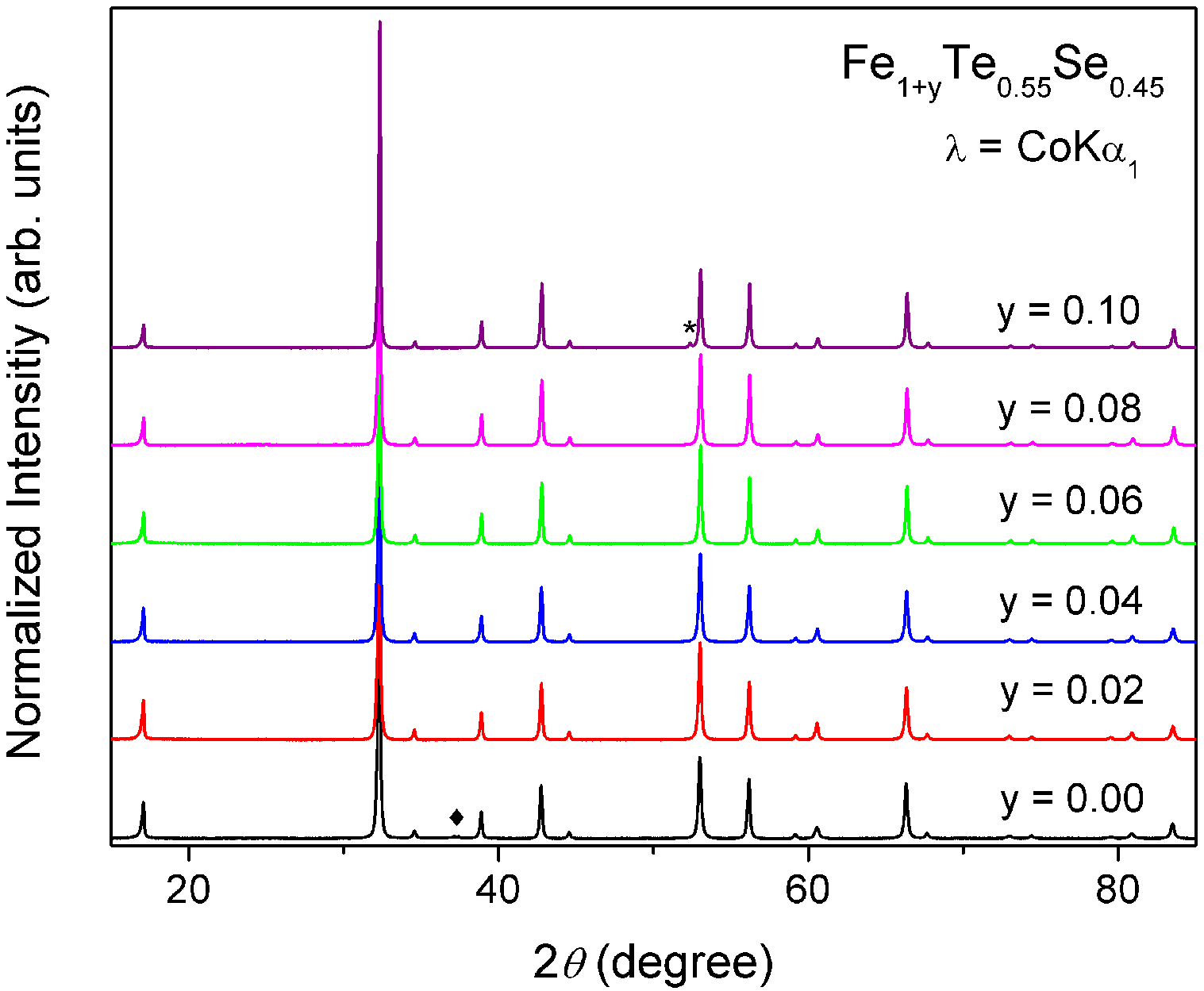}
	\includegraphics[width=0.4\textwidth]{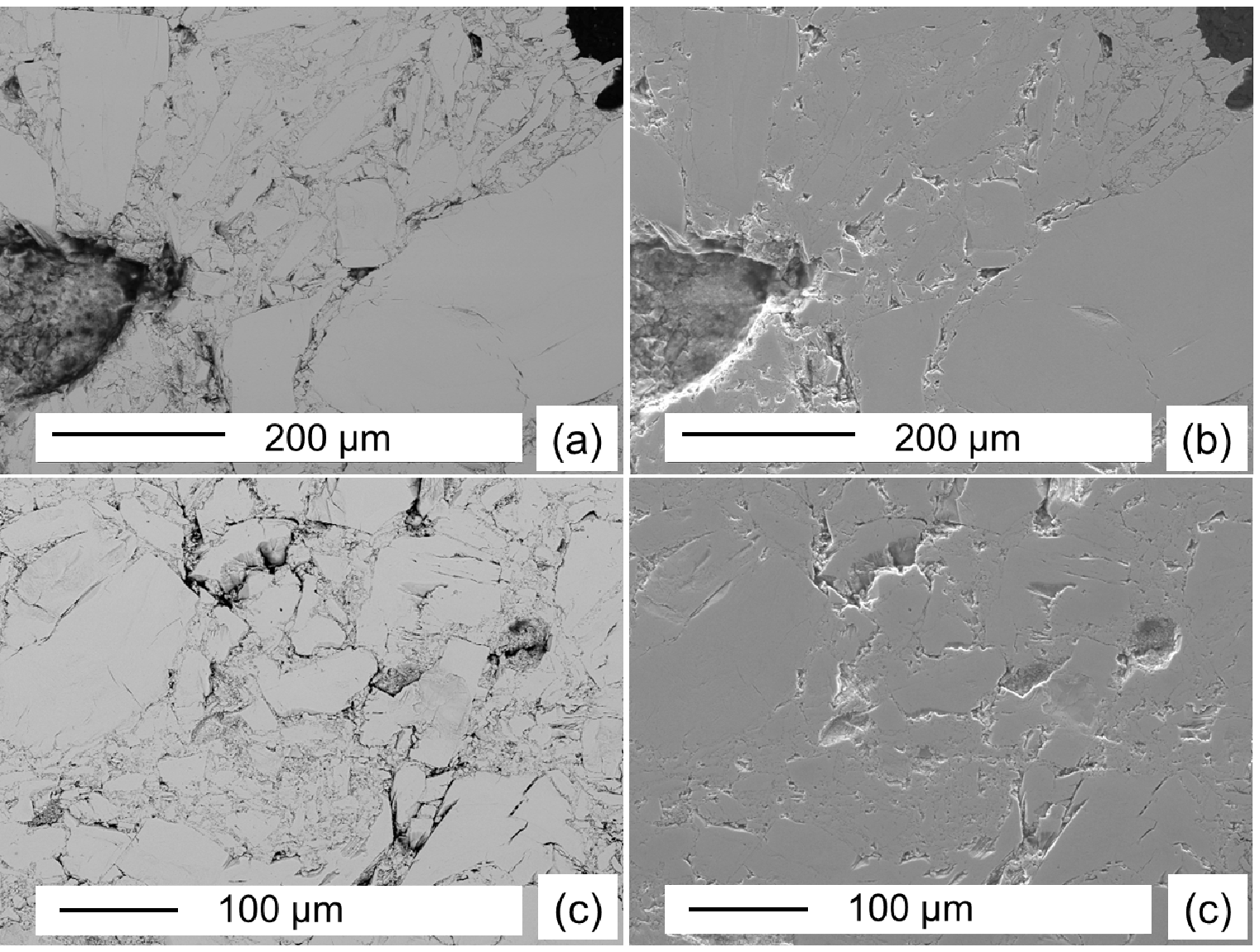}
\caption{Upper panel: PXRD diagram of Fe$_{1+y}$Te$_{0.55}$Se$_{0.45}$ ($y = 0.00-0.10$) annealed at 973 K
for 2 days. $\ast$ and $\blacklozenge$ indicate unreacted Fe and the impurity phase related
to NiAs-type $\delta$-Fe$_{1-y}$Se, respectively. Lower panel: Backscattered (a and c) and secondary electron (b and d) images of annealed
samples Fe$_{1.06}$Te$_{0.55}$Se$_{0.45}$ and Fe$_{1.04}$Te$_{0.55}$Se$_{0.45}$, respectively, indicating single phase material.}
\label{fig:3}       
\end{figure}
\begin{table*}[t]
\caption{Compositions obtained from the EDX analysis of as-grown as well as annealed
polycrystalline samples Fe$_{1+y}$Te$_{0.55}$Se$_{0.45}$. The as-grown samples phase-separated into Phase 1 and Phase 2. After annealing the samples at 973 K for 2 days, single phase materials could be obtained.}
\label{tab:2}       
\centering
\begin{tabular}{llll}
\hline\noalign{\smallskip}
y & Phase 1 & Phase 2&after annealing (single phase)\\  
\noalign{\smallskip}\hline\noalign{\smallskip}
0.00 &Fe$_{1.02(2)}$Te$_{0.56(2)}$Se$_{0.44(2)}$ &Fe$_{0.98(1)}$Te$_{0.35(2)}$Se$_{0.65(2)}$&Fe$_{1.04(1)}$Te$_{0.57(2)}$Se$_{0.43(2)}$ \\
0.02& Fe$_{1.01(1)}$Te$_{0.54(5)}$Se$_{0.46(5)}$ &Fe$_{0.95(1)}$Te$_{0.22(1)}$Se$_{0.78(1)}$&Fe$_{1.06(3)}$Te$_{0.58(1)}$Se$_{0.42(1)}$ \\
0.04 &Fe$_{1.04(1)}$Te$_{0.54(1)}$Se$_{0.46(1)}$&Fe$_{0.95(1)}$Te$_{0.26(6)}$Se$_{0.74(6)}$&Fe$_{1.07(3)}$Te$_{0.58(2)}$Se$_{0.42(2)}$\\
0.06 &Fe$_{1.09(2)}$Te$_{0.56(2)}$Se$_{0.44(2)}$&Fe$_{1.08(1)}$Te$_{0.37(3)}$Se$_{0.63(3)}$&Fe$_{1.10(1)}$Te$_{0.57(1)}$Se$_{0.43(1)}$ \\
0.08 &Fe$_{1.12(3)}$Te$_{0.56(1)}$Se$_{0.44(1)}$& Fe$_{1.06(4)}$Te$_{0.33(1)}$Se$_{0.67(1)}$&Fe$_{1.12(2)}$Te$_{0.57(1)}$Se$_{0.43(1)}$ \\
0.10 &Fe$_{1.13(1)}$Te$_{0.56(1)}$Se$_{0.44(1)}$ &Fe&Fe$_{1.12(1)}$Te$_{0.57(2)}$Se$_{0.43(2)}$\\
\noalign{\smallskip}\hline
\end{tabular}
\end{table*}
\begin{figure}[h]
  \includegraphics[width=0.475\textwidth]{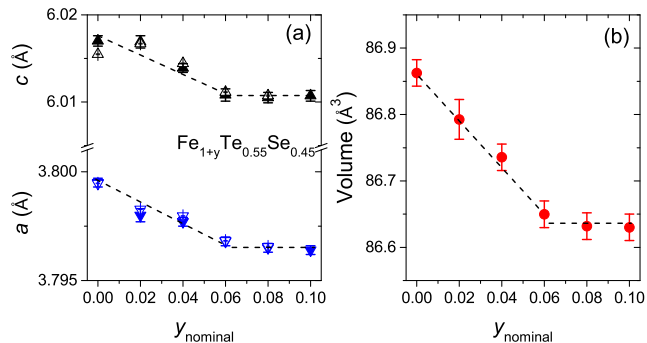}
\caption{(a) Lattice parameters and (b) unit cell volume as a function of 
nominal Fe composition for annealed samples Fe$_{1+y}$Te$_{0.55}$Se$_{0.45}$ ($0.00 \leq y \leq 0.10$).}
\label{fig:4}       
\end{figure}

For the series Fe$_{1+y}$Te$_{0.55}$Se$_{0.45}$, the as-grown samples displayed a chemical phase separation into two ternary phases. The compositions of the two phases are listed in Table \ref{tab:2}. However, after annealing the samples at 973 K for 2 days, chemically homogeneous samples could be obtained. The compositions of the annealed samples are also listed in Table \ref{tab:2} while their PXRD patterns are presented in Fig. \ref{fig:3}, top panel. Impurities were observed only for samples with $y =0$ and $y =0.1$. The bottom panel of Fig. \ref{fig:3} displays back-scattered and secondary electron images of samples with nominal compositions Fe$_{1.06}$Te$_{0.55}$Se$_{0.45}$ and Fe$_{1.04}$Te$_{0.55}$Se$_{0.45}$. These images do not display any secondary phases. Fig. \ref{fig:4} shows the variation of lattice parameters ($a$ and $c$) and unit cell volumes of annealed Fe$_{1+y}$Te$_{0.55}$Se$_{0.45}$ samples. Both lattice parameters and unit cell
volume decrease with increasing Fe-content up to $y = 0.06$. A further increase of the Fe concentration, $i.e.$, ($y \geq 0.08$) does not change the lattice parameters. As a summary of our PXRD,
EDX, and lattice parameter analysis of this series, single phase materials
Fe$_{1+y}$Te$_{0.55}$Se$_{0.45}$ can be obtained when the nominal Fe-content falls into the range $0.00
< y \leq 0.06$. For further increase in Se ($x \geq 0.5$) in Fe$_{1+y}$Te$_{1-x}$Se$_x$, a chemically homogeneous phase could not be obtained even after annealing the samples. Although it is known that long-time annealing of these ternary samples at high temperatures homogenizes the distribution of Se and Te in a crystal, removes local lattice distortions, and induces bulk superconductivity \cite{Taen2009,Noji2010,Noj2012}, single phase samples of  Fe$_{1+y}$Te$_{1-x}$Se$_x$ for $x \geq 0.5$ are not reported in literature.

\begin{figure}[h]
  \includegraphics[width=0.475\textwidth]{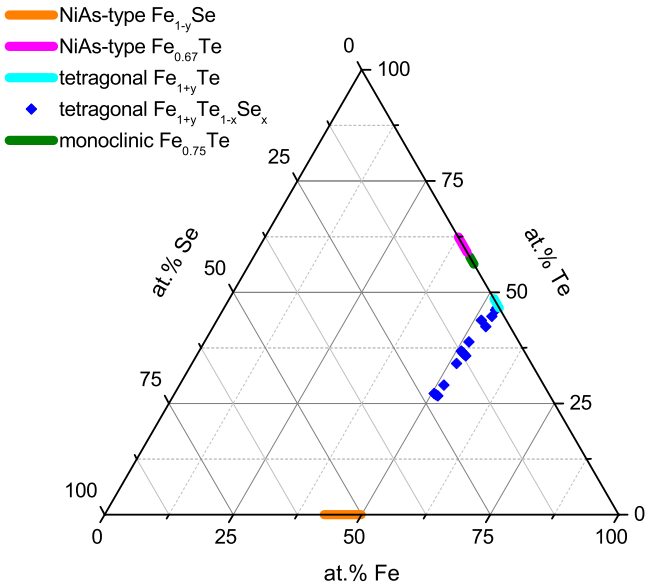}
	\includegraphics[width=0.475\textwidth]{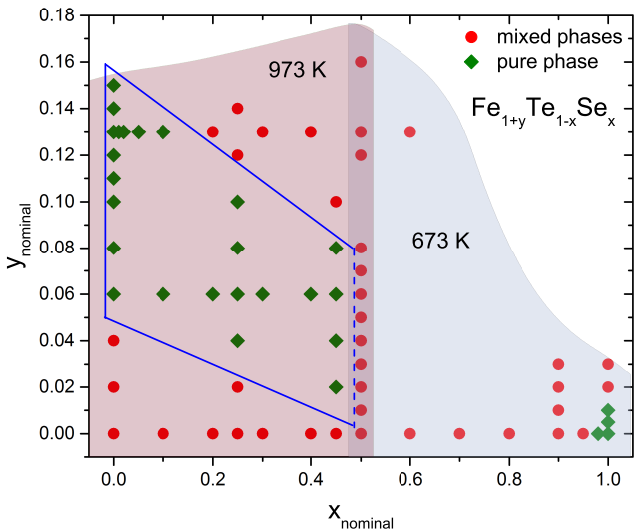}
\caption{Upper panel: Ternary phase diagram of the Fe-Te-Se system. Blue dots indicate single phase of tetragonal Fe$_{1+y}$Te$_{1-x}$Se$_x$. The values for homogeneity
ranges of NiAs-type  $\delta$-Fe$_{1-y}$Se and Fe$_{0.67}$Te, and monoclinic Fe$_{0.75}$Te phases Fe$_{0.75}$Te
are taken from the Pauling File Inorganic Materials Database \cite{Vil2002}. Lower panel: Homogeneity range of Fe$_{1+y}$Te$_{1-x}$Se$_x$ given in a two-dimensional plot for clarity.}
\label{fig:5}       
\end{figure}
Based on our studies, we constructed a ternary phase diagram of the Fe-Te-Se system for homogeneous compositions of Fe$_{1+y}$Te$_{1-x}$Se$_x$, see Fig. \ref{fig:5}. The values of the homogeneity
ranges of NiAs-type $\delta$-Fe$_{1-y}$Se and Fe$_{0.67}$Te, and monoclinic Fe$_{0.75}$Te phases taken from the Pauling File Inorganic Materials Database \cite{Vil2002} are also presented in the upper panel. It can be seen that single phases of tellurium-rich compositions can be obtained in the
presence of excess Fe. For example, compounds of Fe$_{1+y}$Te$_{0.55}$Se$_{0.45}$ and Fe$_{1+y}$Te$_{0.75}$Se$_{0.25}$ can be
realized without impurity phase when the nominal Fe-content falls into the range $0.00
< y \leq 0.08$ and $0.02 < y < 0.12$, respectively. Upon increasing Fe content, the feasible
substitution amount of Se decreases. For $y$ = 0.13, Se substitution is possible in the
range $0.00 \leq x < 0.20$, whereas for $y$ = 0.06 single phase samples of Fe$_{1.06}$Te$_{1-x}$Se$_{x}$ can
be prepared with $0.00 \leq x \leq 0.45$. For low Fe content ($y \approx 0$), impurity peaks of NiAs-type
Fe$_{1-y}$Te$_{1-x}$Se$_{x}$ are observed, whereas for Fe contents $y \geq 0.12$ in Fe$_{1+y}$Te$_{0.75}$Se$_{0.25}$,
elemental Fe remains unreacted.

\begin{figure}[h]
  \includegraphics[width=0.475\textwidth]{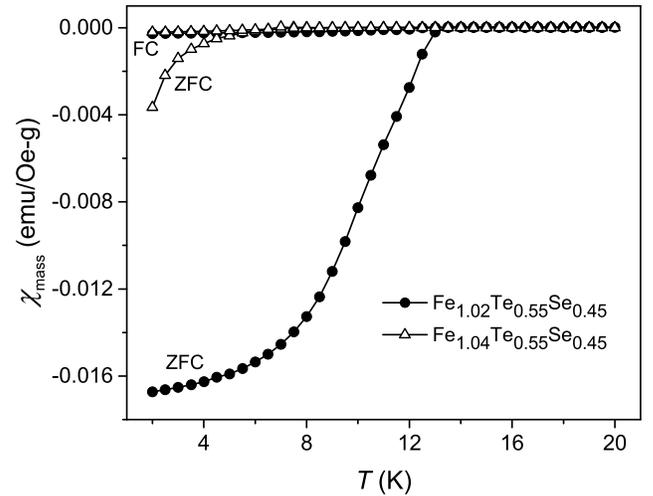}
\caption{Magnetic susceptibility $\chi(T)$ measured in the field cooled (FC) and zero field cooled (ZFC) protocols for Fe$_{1.02}$Te$_{0.55}$Se$_{0.45}$ and Fe$_{1.04}$Te$_{0.55}$Se$_{0.45}$.}
\label{fig:6}       
\end{figure}


In order to investigate the effect of excess Fe on the superconducting properties of Fe$_{1+y}$Te$_{1-x}$Se$_x$, we performed magnetization measurements on phase-pure samples. We find that both the superconducting volume fraction as well as the $T_c$ of the materials drastically decrease when the amount of excess Fe is increased by small amounts. This can clearly be inferred from the example presented in Fig. \ref{fig:6} where magnetic susceptibilities $\chi(T)$ measured in the field cooled (FC) and zero field cooled (ZFC) protocols for Fe$_{1.02}$Te$_{0.55}$Se$_{0.45}$ and Fe$_{1.04}$Te$_{0.55}$Se$_{0.45}$ are compared. For Fe$_{1.02}$Te$_{0.55}$Se$_{0.45}$, the onset of the superconducting transition is $\approx$ 13~K, with a large diamagnetic shielding, which appears to saturate at low temperatures. Upon a 2~$\%$ increase in excess Fe, it can be seen that both the onset of superconductivity and the superconducting shielding factor decreases drastically. A similar behavior of $\chi(T)$ was observed for samples with $y > 0.02$. These studies confirm that, even if the samples are chemically homogeneous, bulk superconductivity occurs for the samples with lowest amount of excess Fe possible. 

\section{Conclusions}
We synthesized a series of ternary compounds with compositions Fe$_{1+y}$Te$_{1-x}$Se$_x$ in order to determine their chemical homogeneity range. For single-phase materials, we found that the maximum amount of excess Fe decreases with increase in Se substitution. For compounds with $x \geq 0.5$, our synthesis procedure did not yield homogeneous compositions. Based on our studies we constructed a ternary phase diagram of the Fe-Te-Se system. We also showed that even in chemically homogeneous compounds, the superconducting volume fraction as well as the transition temperature $T_c$ are rapidly suppressed by an increase in the amount of excess Fe.


\begin{thebibliography}{}

\bibitem{Mc2009}
T. M. McQueen, Q. Huang, V. Ksenofontov, C. Felser, Q. Xu, H. Zandbergen, Y. S. Hor, J.
Allred, A. J. Williams, D. Qu, J. Checkelsky, N. P. Ong, and R. J. Cava, 
Extreme sensitivity of superconductivity to stoichiometry in Fe$_{1+\delta}$Se, 
Phys. Rev. B \textbf{79}, 014522 (2009).

\bibitem{Koz2014}
C. Koz, M. Schmidt, H. Borrmann, U. Burkhardt, S. R\"o{\ss}ler, W. Carrillo-Cabrera, W. Schnelle, U. Schwarz, and Y. Grin, 
Synthesis and Crystal Growth of Tetragonal $\beta$-Fe$_{1.00}$Se,
Z. Anorg. Allg. Chem. \textbf{640}, 1600 (2014).

\bibitem{Rod2011}
E. E. Rodriguez, C. Stock, P. Zajdel, K. L. Krycka, C. F. Majkrzak, P. Zavalij, and M. A. Green, 
Magnetic-crystallographic phase diagram of the superconducting parent compound Fe$_{1+x}$Te,
Phys. Rev. B \textbf{84}, 064403 (2011).

\bibitem{Ros2011}
S. R\"o{\ss}ler, D. Cherian, W. Lorenz, M. Doerr, C. Koz, C. Curfs, Yu. Prots, U. K. R\"o{\ss}ler, U. Schwarz, S. Elizabeth, and S. Wirth, 
First-order structural transition in the magnetically ordered phase of Fe$_{1.13}$Te,
Phys. Rev. B \textbf{84}, 174506 (2011).

\bibitem{Koz2013}
C. Koz, S. R\"o{\ss}ler, A. A. Tsirlin, S. Wirth, and U. Schwarz, 
Low-temperature phase diagram of Fe$_{1+y}$Te studied using x-ray diffraction
Phys. Rev. B \textbf{88}, 094509 (2013).

\bibitem{Rod2013}
E. E. Rodriguez, D. A. Sokolov, C. Stock, M. A. Green, O. Sobolev, J. A. Rodriguez-Rivera, H. Cao, and A. Daoud-Aladine,
Magnetic and structural properties near the Lifshitz point in Fe$_{1+x}$Te,
Phys. Rev. B \textbf{88}, 165110 (2013).

\bibitem{Bao2009}
W. Bao, Y. Qiu, Q. Huang, M. A. Green, P. Zajdel, M. R. Fitzsimmons, M. Zhernenkov, S. Chang, M. Fang, B. Qian, E. K. Vehstedt, J. Yang, H. M. Pham, L. Spinu, and Z. Q. Mao,
Tunable $(\delta\pi , \delta\pi)$-type antiferromagnetic order in $\alpha$ -Fe(Te,Se) superconductors.
Phys. Rev. Lett. \textbf{102}, 247001 (2009).

\bibitem{Yeh2008}
K.-W. Yeh, T.-W. Huang, Y.-L. Huang, T.-K. Chen, F.-C. Hsu, P.M. Wu, Y.-C. Lee, Y.-Y. Chu, C.-L. Chen, J.-Y. Luo, D.-C. Yan, and M.-K. Wu, 
Tellurium substitution effect on superconductivity of the $\alpha$-phase iron selenide,
Europhys. Lett. \textbf{84}, 37002 (2008).

\bibitem{Fang2008}
M. H. Fang, H. M. Pham, B. Qian, T. J. Liu, E. K. Vehstedt, Y. Liu, L. Spinu, and Z. Q. Mao, 
Superconductivity close to magnetic instability in Fe(Se$_{1-x}$Te$_x$)$_{0.82}$,  
Phys. Rev. B \textbf{78}, 224503 (2008).

\bibitem{Liu2009}
T. J. Liu, X. Ke, B. Qian, J. Hu, D. Fobes, E. K. Vehstedt, H. Pham, J. H. Yang, M. H. Fang, L. Spinu, P. Schiffer, Y. Liu, and Z. Q. Mao,
Charge-carrier localization induced by excess Fe in the superconductor Fe$_{1+y}$Te$_{1-x}$Se$_x$,  
Phys. Rev. B \textbf{80}, 174509 (2009). 

\bibitem{Ros2010}
S. R\"o{\ss}ler, D. Cherian, S. Harikrishnan, H. L. Bhat, S. Elizabeth, J. A. Mydosh, L. H. Tjeng, F. Steglich, and S. Wirth, 
Disorder-driven electronic localization and phase separation in superconducting Fe$_{1+y}$Te$_{0.5}$Se$_{0.5}$ single crystals,
Phys. Rev. B \textbf{82}, 144523 (2010).

\bibitem{Che2015}
D. Cherian, S. R\"o{\ss}ler, S. Wirth, and S. Elizabeth,
Interplay of structure, magnetism, and superconductivity in Se substituted iron telluride with excess Fe,
J. Phys.: Condens. Matter, \textbf{27}, 205702 (2015).

\bibitem{Gro1954}
F. Gr\o nvold, H. Haraldsen and J. Vihovde,
Phase and structural relations in the system iron tellurium, 
Acta Chem. Scand. \textbf{8}, 1927 (1954).

\bibitem{Ter1981}
P. Terzieff, 
The magnetism of the NiAs-type solid solution Fe-Se-Te,
Physica B+C \textbf{103}, 158 (1981).

\bibitem{Taen2009}
T. Taen, Y. Tsuchiya, Y. Nakajima, and T. Tamegai,
Superconductivity at $T_c= 14$~K  in single-crystalline FeTe$_{0.61}$Se$_{0.39}$,  
Phys. Rev. B, 80, 092502 (2009).

\bibitem{Noji2010}
T. Noji, T. Suzuki, H. Abe, T. Adachi, M. Kato, and Y. Koike,
Growth, Annealing Effects on Superconducting and Magnetic Properties, and Anisotropy of FeSe$_{1-x}$Te$_x$ ($0.5\leq x \leq 1$) Single Crystals,
J. Phys. Soc. Jpn. \textbf{79}, 084711 (2010).

%
\bibitem{Noj2012}
T. Noji,  M. Imaizumi,  T. Suzuki,  T. Adachi,  M. Kato, and  Y. Koike,
Specific-heat study of superconducting and normal states in FeSe$_{1-x}$Te$_x$ ($0.6\leq x \leq 1$) single crystals: strong-coupling superconductivity, strong electron-correlation, and inhomogeneity,
J. Phys. Soc. Jpn. \textbf{81}, 054708 (2012).

\bibitem{Vil2002}
P. Villars, M. Berndt, K. Brandenburg, K. Cenzual, J. Daams, F. Hulliger, T.
Massalski, H. Okamoto, K. Osaki, A. Prince, H. Putz, S. Iwata, PAULING FILE,
Binaries Edition, ASM International, Materials Park, Ohio, (2002).


%
\end{thebibliography}

\section{acknowledgement}
We thank U. Burkhardt and G. Auffermann for their help in sample characterization. We are grateful to Yuri Grin and Liu Hao Tjeng for helpful discussions.
Financial support from the Deutsche Forschungsgemeinschaft within the priority program SPP1458 is gratefully acknowledged.  
\balance

\end{document}